\documentclass[twocolumn]{aastex631}

\usepackage{amsmath}
\usepackage{hyperref}

\newcommand{\Gravitas}{\textsc{Gravitas}}
\newcommand{\Version}{1.0.0}
\newcommand{\ScenarioCount}{59}
\newcommand{\InvestigationCount}{22}
\newcommand{\InvestigationSteps}{636}
\newcommand{\PhysicsChecks}{243}
\newcommand{\ValidationAreas}{25}
\newcommand{\ValidationSources}{50}
\newcommand{\ReleaseDOI}{10.5281/zenodo.22800610}
\newcommand{\ZenodoRecord}{https://zenodo.org/records/22800610}
\newcommand{\CodeSnapshot}{9148f9e}
\newcommand{\TableCell}[2]{\parbox[t]{#1}{\raggedright #2}}

\shorttitle{Gravitas}
\shortauthors{Ziegler}

\begin{document}

\title{Gravitas: A Browser-Based Astrophysics Laboratory for Prediction,
Measurement, and Discovery}

\author[0000-0002-0619-7639]{Carl Ziegler}
\affiliation{Department of Physics, Engineering and Astronomy, Stephen F. Austin
State University, Nacogdoches, TX 75962, USA}
\email{Carl.Ziegler@sfasu.edu}

\begin{abstract}
Astronomy students are often asked to reason about systems that cannot be
handled in a laboratory, evolve over inaccessible timescales, or reveal
themselves only through indirect measurements. \Gravitas\ is a free,
browser-based astrophysics environment designed to turn those limitations into
opportunities for experimentation. Its version \Version\ release
combines a configurable gravitational sandbox with \ScenarioCount\ ready-to-run
scenarios and \InvestigationCount\ guided investigations comprising
\InvestigationSteps\ steps and roughly 15--19 hours of structured work. Students can build and
alter systems, change reference frames, measure distances, angles, periods,
energies, forces, and orbital elements, and observe simulated systems through
transit, radial-velocity, astrometric, rotation-curve, stellar-evolution, and
gravitational-wave displays. Guided activities use a
predict--test--measure--revise--explain cycle, retain the student's evidence, and
export a report for submission. Instructor guides, answer keys, short classroom
routes, lecture and embed modes, reproducible share links, Spanish localization,
keyboard-accessible navigation, and offline operation lower the practical
barrier to adopting the simulation in introductory astronomy and physics
courses. A public dashboard
reports \PhysicsChecks\ scientific checks across \ValidationAreas\ areas and names
the sources and tolerances behind them. This article
describes the learning experiences available in \Gravitas, illustrates several
ways it can be used from a brief demonstration to a full laboratory exercise,
and states the scientific scope and limitations of the models. The aim is not to
replace quantitative calculation or observation, but to give students a place
where they can make a prediction, change one thing, measure what follows, and
build an explanation from their own evidence.
\end{abstract}


\section{Introduction}
\label{sec:introduction}

Astronomy is unusually rich in phenomena that students can see but cannot
directly manipulate. They cannot move a planet closer to its star, turn off a
galaxy's dark matter, replay a close stellar encounter, or wait through a
main-sequence lifetime. Even familiar ideas such as an elliptical orbit or a
transit light curve require students to coordinate several representations at
once: a moving physical system, a graph, a mathematical relationship, and the
viewpoint of an observer.

Interactive simulations can make such systems available for inquiry. Studies
and reviews show that well-designed computer simulations can support conceptual
learning in science education and, for a targeted learning goal, can sometimes
outperform physical apparatus \citep{finkelstein2005,wieman2008,rutten2012}.
The broader active-learning
literature likewise reports stronger conceptual understanding and course
performance under interactive-engagement approaches than under traditional
instruction \citep{hake1998,freeman2014}.
\Gravitas\ was built around that distinction. It can be opened as a free-form
sandbox, but it also contains a complete instructional layer that asks students
to predict an outcome before running a system, measure the result with
task-specific instruments, compare it with their prediction, and explain what
changed.

The intended audience is broad: instructors teaching introductory astronomy or
physics, students exploring independently, and presenters who want a
reproducible live demonstration. A visitor needs no account, installation, or
specialized hardware. The simulation runs in a modern web browser at
\url{https://gravitas-sim.online}; the source is openly available under the MIT
License. A complete state can be encoded in a URL, allowing an instructor to
distribute the same initial conditions to an entire class and allowing a student
to return a modified system as part of an answer.

This paper presents \Gravitas\ as a teaching environment rather than as a
software architecture. Section~\ref{sec:experience} introduces the student
experience. Sections~\ref{sec:investigations} and \ref{sec:tools} describe the
guided investigations and measurement tools. Section~\ref{sec:examples} follows
three representative learning experiences, while
Section~\ref{sec:instructors} outlines practical classroom use.
Section~\ref{sec:scope} states what the models can and cannot support. The paper
describes the released version represented by the \texttt{v\Version} tag at
source commit \texttt{\CodeSnapshot}. The corresponding source archive is
preserved by Zenodo at DOI \doi{\ReleaseDOI}.

\section{From First Click to Experiment}
\label{sec:experience}

\Gravitas\ opens directly into an interactive system. The central canvas shows
the evolving scene, while controls expose bodies, scenario settings, time,
viewpoint, and measurement tools. A learner can start from the Solar System,
TRAPPIST-1, a binary star, a spacecraft transfer, a galactic rotation curve, or
a black-hole encounter, then change the masses, positions, velocities, and
physical options that define the experiment. The current gallery contains
\ScenarioCount\ systems spanning orbital motion, exoplanets, stars, galaxies,
compact objects, and playful counterfactual cases. Concept tags let an
instructor find a system that matches the topic of a particular class meeting.

The interface is designed to move naturally between qualitative observation and
quantitative evidence. Trails reveal orbital shape and motion. Velocity and
acceleration arrows expose the changing geometry of force and motion. A
selection inspector reports mass, radius, speed, orbital elements, and energy
for an individual body. A ruler, protractor, stopwatch, plots, and event-aware
measurements let students attach numbers to what they see. The elapsed
simulation time and spatial scale remain visible, and exported screenshots
carry the scenario name, clock, and scale so that an image remains interpretable
after it leaves the application.

\begin{figure*}
\centering
\includegraphics[width=0.96\textwidth]{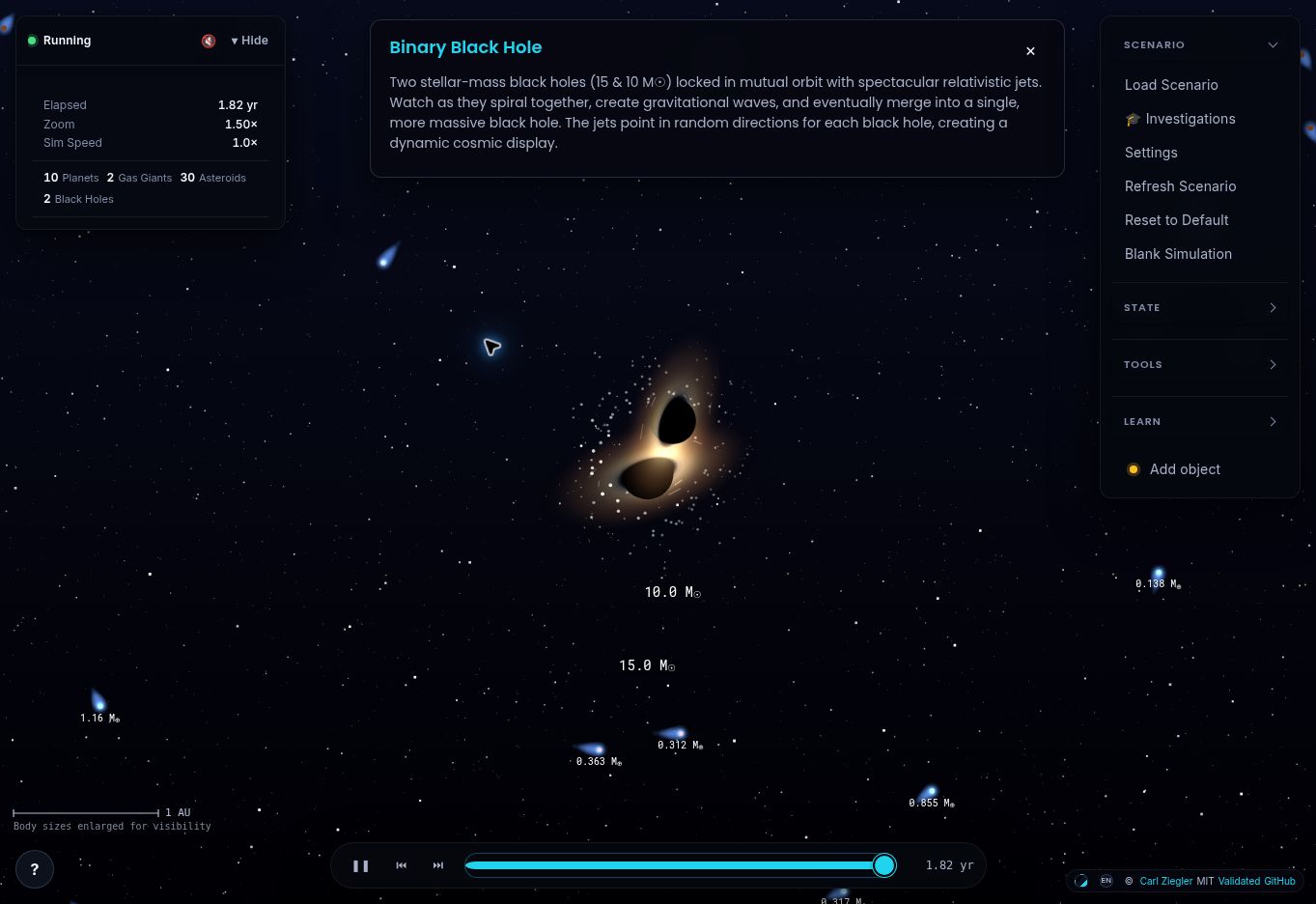}
\caption{The live \Gravitas\ workspace in the Binary Black Hole scenario. The
canvas remains the visual focus, while a persistent readout, simulated-time
timeline, scale bar, scenario description, and collapsible controls keep the
state of the experiment visible. The same interface can be opened without an
account or installation at \url{https://gravitas-sim.online}.}
\label{fig:overview}
\end{figure*}

A scenario is not a locked animation. Students can pause, step, rewind by
restoring a captured state, or change the rate at which simulated time passes.
They can follow a body with the camera or re-express the entire recorded scene
in that body's reference frame. The latter distinction matters: changing to
Earth's frame redraws the prior trails and causes Mars to trace the familiar
retrograde loop, while the underlying heliocentric dynamics remain unchanged.
The student therefore sees reference frame as a choice of description rather
than a change in physical law.

The same emphasis on reproducibility appears throughout the application.
Shareable links preserve the initial configuration. Seeded scenarios reproduce
the same controlled trial. An A/B experiment bench captures a state, records a
baseline, restores the identical state, and then records a second run after one
variable is changed. The resulting comparison names the changed parameter and
can be exported with provenance. These features support a simple but important
classroom habit: change one thing, preserve everything else, and decide whether
the evidence supports the claimed cause.

\begin{figure*}[t]
\centering
\includegraphics[width=0.96\textwidth]{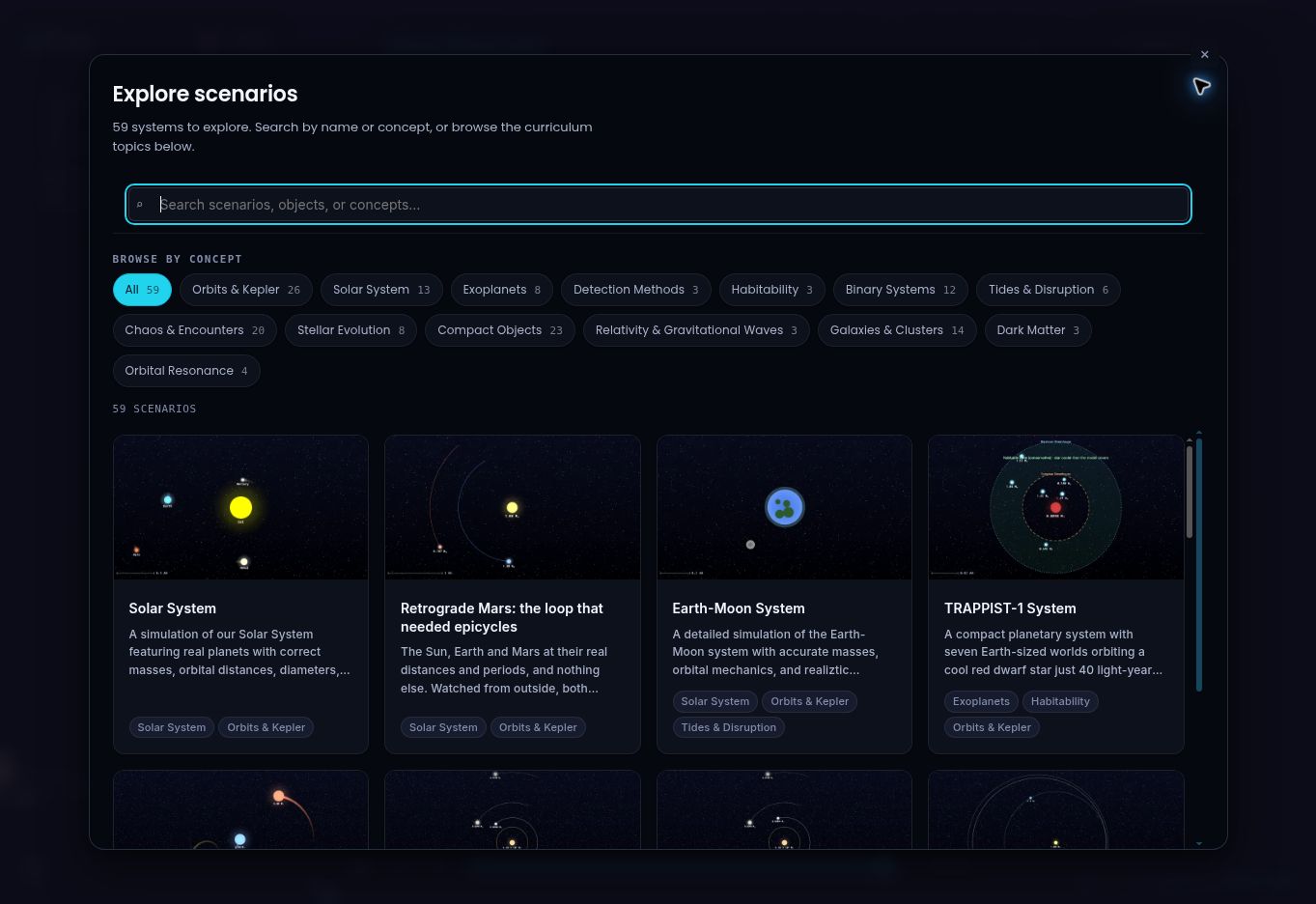}
\caption{The searchable scenario library. Its \ScenarioCount\ systems are
indexed by curriculum concepts including orbits, the Solar System, exoplanets,
stellar evolution, compact objects, galaxies, dark matter, tides, chaos,
resonance, and gravitational waves. Preview images and short descriptions let
an instructor identify a relevant demonstration without first learning the
control system.}
\label{fig:scenarios}
\end{figure*}

\section{Guided Investigations}
\label{sec:investigations}

The \InvestigationCount\ guided investigations contain
\InvestigationSteps\ student-facing steps. They range from short,
15--25 minute spaceflight activities to 80--100 minute stellar-evolution
sequences. All are aimed primarily at introductory astronomy, with one
gravitational-wave investigation explicitly requiring no prior physics.

Each investigation follows a recurring learning cycle:

\begin{enumerate}
\item \textbf{Predict.} The learner commits to an expected outcome before the
simulation reveals it. Predictions are recorded but are not graded for
correctness.
\item \textbf{Test.} A prepared scene loads with the controls and instruments
needed for the question.
\item \textbf{Measure.} The learner records values from the running model rather
than copying numbers from explanatory text.
\item \textbf{Revise.} The learner compares the result with the original
prediction and, where appropriate, changes a variable or repeats a controlled
run.
\item \textbf{Explain.} A short response, calculation, graph, or comparison asks
the learner to connect the evidence to the physical idea.
\end{enumerate}

This structure is intended to limit passive watching. The prediction gives the
result something to challenge; the measurement makes the conclusion auditable;
and the final explanation requires the student to say why the result occurred.
Progress is stored locally, and the finished evidence notebook can be exported
as a PDF for an existing learning-management system. No student account or
external data service is required.

\begin{figure*}[t]
\centering
\includegraphics[width=0.94\textwidth]{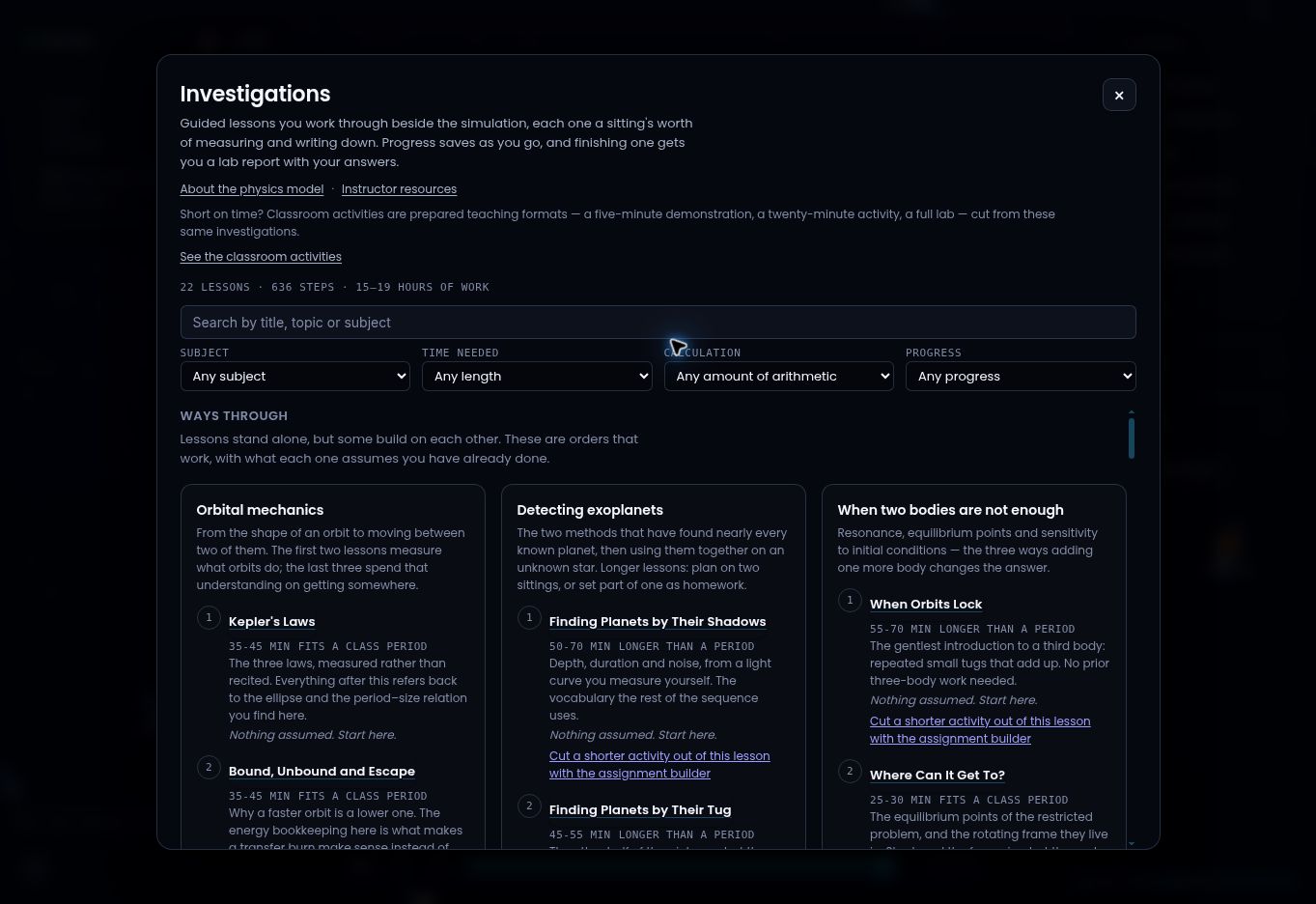}
\caption{The investigation browser exposes the intended time, calculation
load, prerequisites, sequence, and expected student product before a lesson is
opened. Prepared pathways organize lessons into orbital mechanics, exoplanet
detection, three-body dynamics, and gravitational-wave sequences, while
filters help instructors match an activity to a class period.}
\label{fig:investigation-library}
\end{figure*}

\begin{figure*}[t]
\centering
\includegraphics[width=0.96\textwidth]{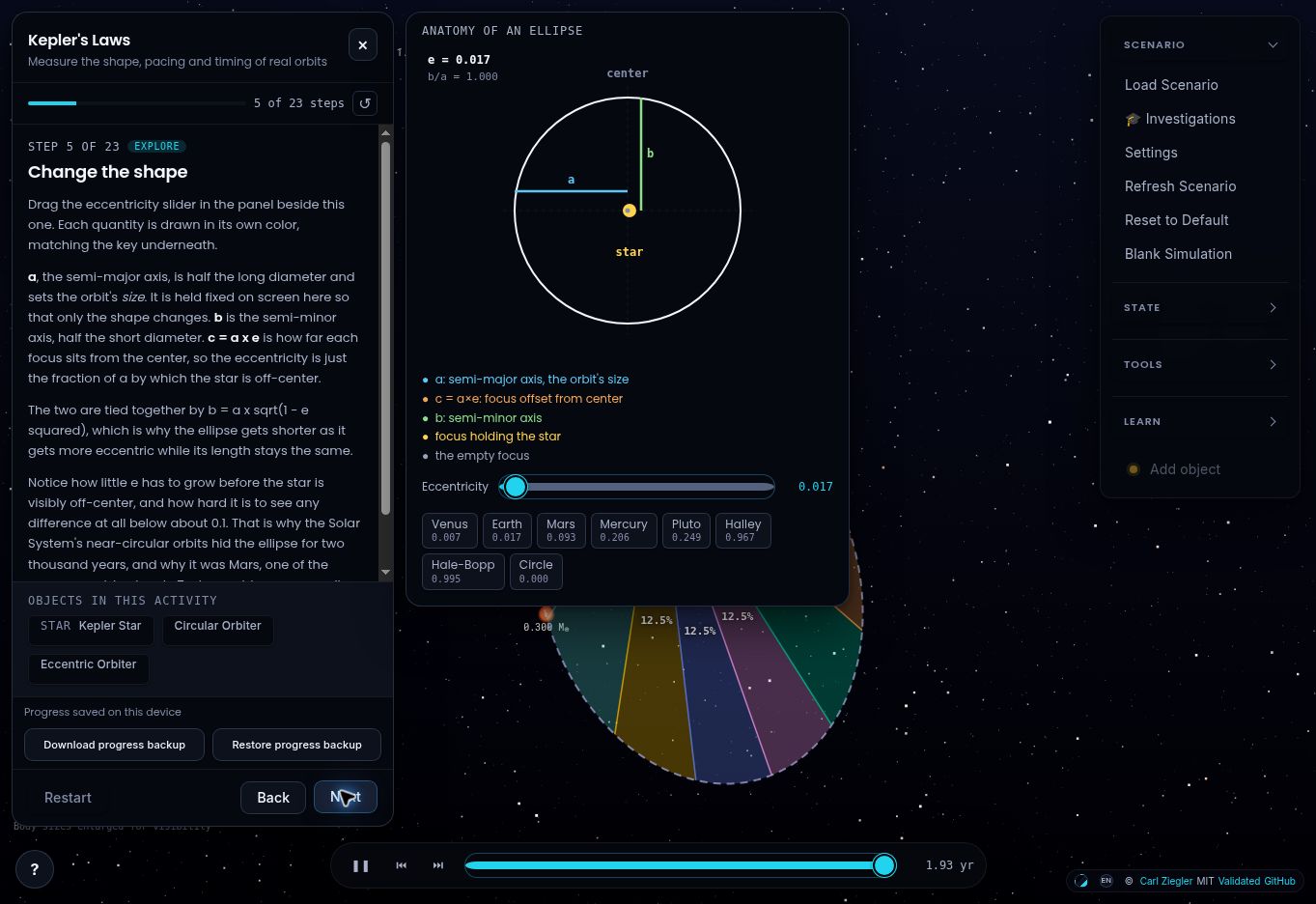}
\caption{A guided step from the Kepler's Laws investigation. The lesson prompt
and saved progress remain beside the running system, while an interactive
ellipse explorer connects eccentricity to the semimajor axis, semiminor axis,
and focus offset. Preset Solar System values invite rapid comparison without
removing the student's control of the experiment.}
\label{fig:workflow}
\end{figure*}

Table~\ref{tab:investigations} summarizes the breadth of the catalogue. The
investigations are not merely tours of features: each is organized around a
claim that students can settle with a measurement. For example, the resonance
activity first demonstrates that a near-integer period ratio is not sufficient
evidence for resonance, then asks students to distinguish libration from
circulation of the resonant angle. The binary-planet activity maps the boundary
between surviving and disrupted orbits. The gravitational-wave sequence asks
students to distinguish what is physically calculated, what is an illustrative
visualization, and what an observer measures.

\begin{table*}[t]
\caption{Guided investigation catalogue}
\label{tab:investigations}
\centering
\small
\renewcommand{\arraystretch}{1.12}
\begin{tabular}{lll}
\hline\hline
\TableCell{0.17\textwidth}{\textbf{Theme}} &
\TableCell{0.31\textwidth}{\textbf{Investigations}} &
\TableCell{0.45\textwidth}{\textbf{Representative student work}}\\
\hline
\TableCell{0.17\textwidth}{Orbits and gravity} &
\TableCell{0.31\textwidth}{
Kepler's Laws; Why Mars Goes Backwards; Bound, Unbound and Escape; Weighing the
Stars; Tides} &
\TableCell{0.45\textwidth}{
Locate the focus of an ellipse; compare equal swept areas; transform reference
frames; classify trajectories by energy; infer stellar mass from a binary
orbit; measure differential gravity.}\\
\TableCell{0.17\textwidth}{Exoplanets and observing} &
\TableCell{0.31\textwidth}{
Finding Planets by Their Shadows; Finding Planets by Their Tug; The Goldilocks
Question; Can You Detect This Planet?; Design the Schedule; Planets in Binary
Stars} &
\TableCell{0.45\textwidth}{
Recover radius and period from a transit; infer mass from radial velocity;
combine mass and radius; explore inclination and dilution; design observations;
map stable regions around one or two stars.}\\
\TableCell{0.17\textwidth}{Orbital dynamics and spaceflight} &
\TableCell{0.31\textwidth}{
The Butterfly Effect in Space; When Orbits Lock; Where Does a Gravity Assist Get
Its Speed?; Getting There From Here; Where Can It Get To?} &
\TableCell{0.45\textwidth}{
Measure divergence of nearby starts; test resonant angles; compare a flyby in
two frames; calculate and fly a Hohmann transfer; use the Jacobi constant and
zero-velocity regions.}\\
\TableCell{0.17\textwidth}{Compact objects and gravitational waves} &
\TableCell{0.31\textwidth}{
Black Holes by the Numbers; What Is a Gravitational Wave?; Listening to
Spacetime} &
\TableCell{0.45\textwidth}{
Test how black-hole properties scale with mass; connect orbital motion to a
travelling strain; measure a chirp and compare the model with GW150914
\citep{abbott2016}.}\\
\TableCell{0.17\textwidth}{Stars, galaxies, and dark matter} &
\TableCell{0.31\textwidth}{A Universe of Stars; Lives of Stars; The Missing Mass} &
\TableCell{0.45\textwidth}{
Separate mass, radius, temperature, and luminosity on the H--R diagram; follow
published stellar tracks \citep{dotter2016,choi2016}; fit a rotation curve and
compare luminous and dynamical mass.}\\
\hline
\end{tabular}
\vspace{3pt}

\parbox{0.95\textwidth}{\footnotesize\textit{Note.} Durations and prerequisites
are supplied in the application. The catalogue can be filtered by concept, and
individual investigations may be assigned through reproducible links.}
\end{table*}

\section{Instructional Design: Freedom with Scaffolding}
\label{sec:design}

The open sandbox and the guided investigations are deliberately part of the
same environment. A student can begin with a tightly framed task, acquire a
measurement habit, and then ask a question that was not anticipated by the
lesson. Conversely, an instructor can begin with an arresting free-form
demonstration and then move directly into a structured activity using the same
controls and the same physical model. The transition from being shown a result
to investigating it does not require another application.

\subsection{Predictions create a reason to measure}

In a guided investigation, the learner records a prediction before the relevant
motion or graph is revealed. The prediction is preserved as evidence of prior
reasoning rather than graded as a right-or-wrong gate. That choice matters: a
wrong prediction can become the most useful part of the activity if the student
can identify the observation that changed it. The next steps therefore ask for
a measured quantity or comparison, followed by a revision and explanation.

This structure is consistent with the wider evidence that interactive
engagement and well-scaffolded simulations are more effective than passive
exposure alone \citep{hake1998,wieman2008,rutten2012,freeman2014}. \Gravitas\
does not assume that movement on a screen is itself active learning. Activity
is organized around a decision: what to vary, what to hold fixed, what to
measure, and what conclusion the measurement permits.

\subsection{Several representations remain connected}

Many astronomy misconceptions arise when a student treats a diagram, a graph,
and an equation as unrelated facts. In \Gravitas, these representations are
driven by one evolving state. Moving the observer changes the apparent orbit,
the transit light curve, the radial-velocity signal, and the astrometric track
together. Changing an orbit changes both the moving scene and the point that
appears on a period--semimajor-axis plot. Switching a dark halo off changes the
accelerations of the stars and the rotation curve used to explain them. The
student is invited to move back and forth between picture, plot, and numerical
readout until the representations tell one coherent story.

The program also distinguishes quantities that are calculated from quantities
that are only drawn. A body may be enlarged on the canvas for visibility, but
its radius is read from the inspector or the attached instrument. Decorative
jets and an illustrative spacetime surface are identified as visualizations,
while the velocities and accelerations used in an orbit come from the running
model. This distinction gives instructors a natural opening for discussing what
counts as evidence in a simulation.

\subsection{Evidence survives the moment of discovery}

The evidence notebook gathers predictions, measurements, plots, screenshots,
and written responses as the student works. A finished investigation can be
exported as a PDF, while numerical time series can be exported as unit-labelled
CSV files. The report is therefore not a separate worksheet that happens to sit
next to a simulation; it is a record made from the same states and instruments
that produced the observation.

This design supports assessment without turning every interaction into a quiz.
An instructor can evaluate whether the student stated a testable prediction,
changed a defensible variable, recorded an appropriate measurement, interpreted
the uncertainty or limitation, and connected the result to the physical idea.
Those criteria work for a five-paragraph report, a brief in-class response, or
an oral explanation built around a shared link. They also make a student's
reasoning visible when the numerical answer alone would hide it.

\section{Tools That Turn Motion into Evidence}
\label{sec:tools}

The tools in \Gravitas\ are selected for questions commonly asked in
introductory astronomy. They are available in the sandbox and are also opened
automatically when a guided step requires them.

\begin{table*}[t]
\caption{Student tools and the reasoning they support}
\label{tab:tools}
\centering
\small
\renewcommand{\arraystretch}{1.12}
\begin{tabular}{lll}
\hline\hline
\TableCell{0.21\textwidth}{\textbf{Tool}} &
\TableCell{0.34\textwidth}{\textbf{What the student records}} &
\TableCell{0.38\textwidth}{\textbf{Example question}}\\
\hline
\TableCell{0.21\textwidth}{Ruler, protractor, and scale bar} &
\TableCell{0.34\textwidth}{
Distances in astronomical units or kilometres; angles between locations or
vectors} &
\TableCell{0.38\textwidth}{
Where is the focus? How far does a planet move in a fixed interval? What angle
separates a Trojan asteroid from Jupiter?}\\
\TableCell{0.21\textwidth}{Simulation-time stopwatch} &
\TableCell{0.34\textwidth}{Elapsed time, including automatic latching at successive periapsis passages} &
\TableCell{0.38\textwidth}{How can a period be measured without relying on reaction time?}\\
\TableCell{0.21\textwidth}{Object inspector} &
\TableCell{0.34\textwidth}{Mass, radius, speed, orbital elements, and kinetic, potential, and total energy} &
\TableCell{0.38\textwidth}{Is the trajectory bound? Which body is the orbit actually referenced to?}\\
\TableCell{0.21\textwidth}{Velocity, acceleration, and source arrows} &
\TableCell{0.34\textwidth}{
The selected body's velocity, total acceleration, and contribution from each
gravitating source} &
\TableCell{0.38\textwidth}{Why is an orbit curved? Why are velocity and force not generally parallel?}\\
\TableCell{0.21\textwidth}{Reference-frame controls} &
\TableCell{0.34\textwidth}{
The same positions and recorded trails relative to the barycenter or a selected
body} &
\TableCell{0.38\textwidth}{Why does Mars reverse direction in an Earth-centered view?}\\
\TableCell{0.21\textwidth}{A/B experiment bench} &
\TableCell{0.34\textwidth}{
Aligned baseline and comparison time series, changed parameter, CSV, and
provenance metadata} &
\TableCell{0.38\textwidth}{
Does changing the starting distance, rather than the random seed or timestep,
cause the different outcome?}\\
\TableCell{0.21\textwidth}{Observing panels} &
\TableCell{0.34\textwidth}{
Transit flux, radial velocity, astrometric displacement, viewing geometry, and
sampled observations} &
\TableCell{0.38\textwidth}{
Which properties of a planet can each method recover, and how does inclination
change the evidence?}\\
\TableCell{0.21\textwidth}{Specialized plots} &
\TableCell{0.34\textwidth}{
Swept area, rotation curve, resonant angle, stellar tracks, gravitational-wave
strain, and conservation diagnostics} &
\TableCell{0.38\textwidth}{
What observation distinguishes a resonance from a near-integer ratio? Why does
a galaxy require more gravitating mass than its light implies?}\\
\TableCell{0.21\textwidth}{Exports} &
\TableCell{0.34\textwidth}{Screenshots, clips, report PDF, state link, and unit-labelled CSV} &
\TableCell{0.38\textwidth}{Can another student or instructor reproduce the setup and inspect the evidence?}\\
\hline
\end{tabular}
\end{table*}

Three observing panels are especially useful for connecting dynamics to
astronomical inference. A transit light curve, a radial-velocity curve, and an
astrometric track are generated from the same evolving system and the same
observer geometry. Moving the observer therefore changes all three consistently:
a planet can cease to transit, the radial-velocity amplitude changes with
inclination, and the astrometric line opens into an ellipse. Students encounter
selection effects and degeneracies as outcomes of a controlled experiment,
rather than only as warnings appended to a formula.

\begin{figure*}
\centering
\includegraphics[width=0.94\textwidth]{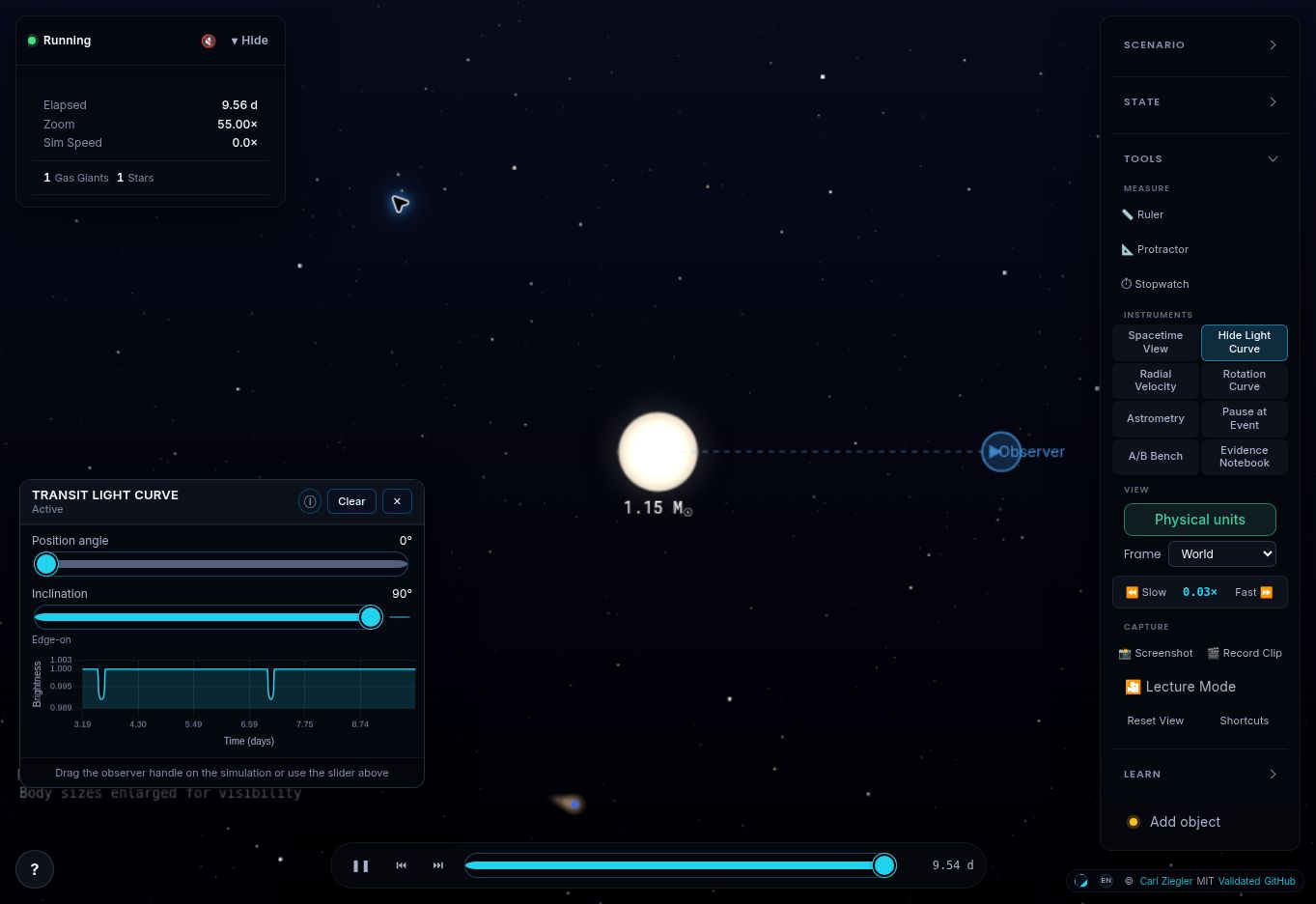}
\caption{Transit photometry in the Exoplanet Characterization Lab. The plot is
generated from the live orbit, and the observer handle on the canvas shares its
position angle and inclination with every observing panel. Moving away from an
edge-on view removes the transit and makes the geometric selection effect
immediately visible.}
\label{fig:exoplanets}
\end{figure*}

\begin{figure*}[t]
\centering
\includegraphics[width=0.94\textwidth]{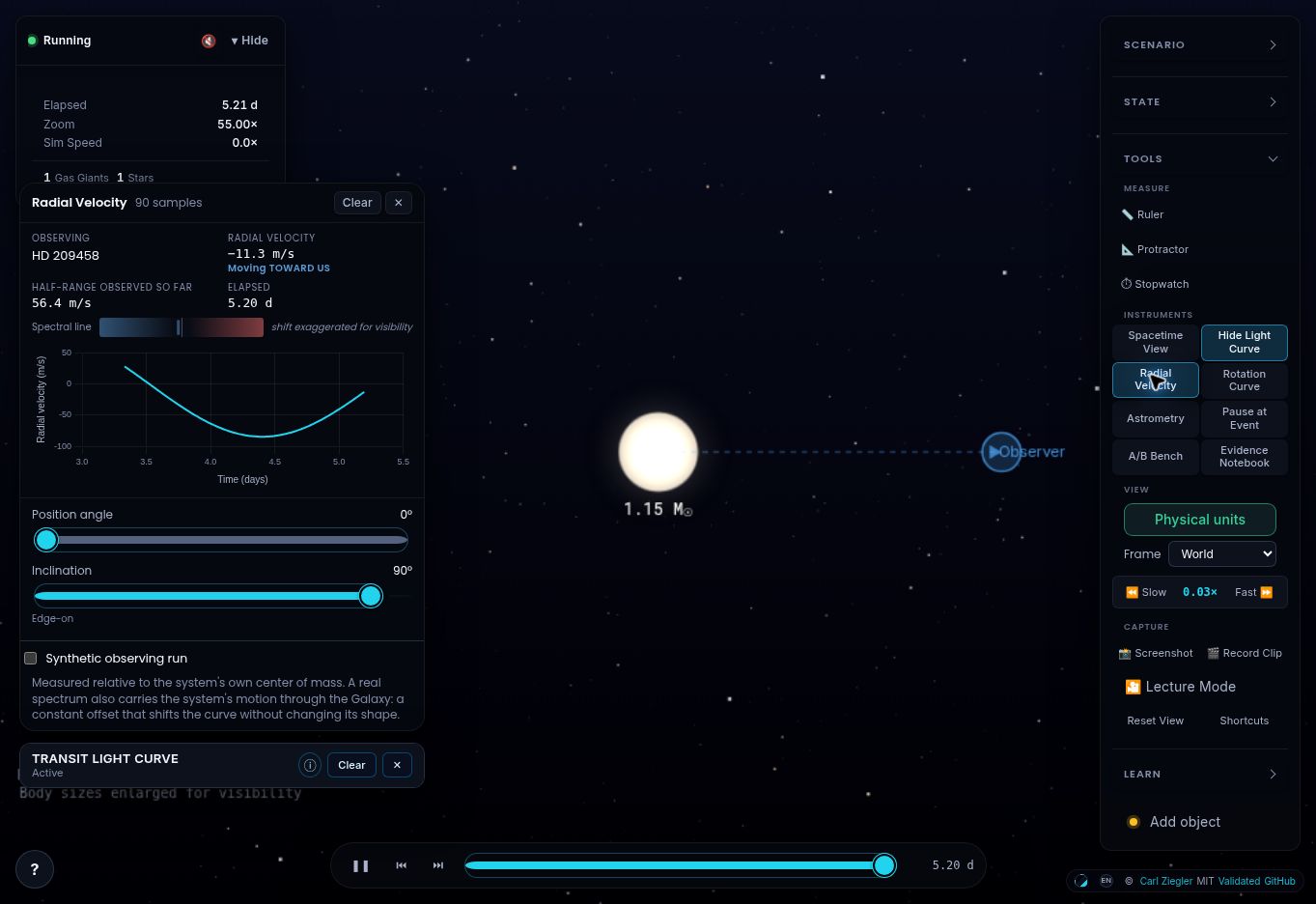}
\caption{The same system observed through stellar radial velocity. The panel
shows the growing time series, the instantaneous sign and amplitude of the
motion, the displaced spectral line, and the shared viewing controls. Students
can compare what transit and radial-velocity measurements reveal before
combining them to infer the planet's bulk density.}
\label{fig:radial-velocity}
\end{figure*}

The application also makes numerical reliability visible when it is relevant.
Students may choose among three integrators and inspect changes in total energy
and angular momentum. Dedicated investigations distinguish a physical
sensitivity to initial conditions from ordinary phase drift or numerical error.
The purpose is not to teach numerical analysis in every activity; it is to let
students ask whether a surprising result survives an appropriate reliability
check.

\section{Three Paths Through the Simulation}
\label{sec:examples}

\subsection{Discovering Kepler's Laws}

In the Kepler investigation, students begin with the geometry of an ellipse.
They locate the star relative to the two foci, alter eccentricity, and compare
circular and eccentric orbits. The simulation then displays equal-area sectors
over equal time intervals while the student independently measures the fastest
and slowest portions of the orbit. Finally, periods and semimajor axes from
multiple systems are plotted to recover a slope of $3/2$ in log space.

This sequence connects picture, measurement, and equation. The second law is not
reduced to the statement that a planet moves faster when it is closer. The
learner sees the swept area, times the motion, and relates the result to angular
momentum. The third law is not supplied as an exponent to verify; it is the
relationship that emerges from the student's graph. The same content can be
used as a five-minute prediction and demonstration, a roughly twenty-minute
guided activity, or a full laboratory investigation.

\subsection{Becoming an Exoplanet Observer}

The exoplanet sequence begins with the shadow of HD~209458~b, a landmark
transiting system \citep{charbonneau2000}. Students measure a transit depth,
infer a radius, examine limb darkening, determine a period, and explore how
blended light from a companion star biases the result. The radial-velocity
investigation then uses the star's reflex motion to estimate the planet's mass.
Combining the two measurements produces a bulk density and a more physically
meaningful description of the planet.

Later investigations shift the question from parameter recovery to experimental
design. Students receive a finite number of observing nights, choose a cadence,
and compare schedules under the same signal and noise realization. They see how
aliasing, weather losses, baseline, and sampling affect whether a planet is
detected. The Goldilocks investigation then changes orbital distance, stellar
luminosity, and eccentricity to examine incident flux and the limits of the
phrase habitable zone, using established habitable-zone relations as the
scientific context \citep{kopparapu2013}.

Together, these activities reproduce the logic of observation: no single panel
reveals everything, geometry matters, and the observing strategy can determine
which conclusions are possible.

\subsection{When the Visible Matter Is Not Enough}

The Missing Mass investigation asks students to weigh systems twice. First they
sum the matter that is visible; then they infer gravitating mass from orbital
motion. These estimates agree in a Solar System-like case but diverge for a
galaxy and a galaxy cluster. Students arrange visible mass, inspect the rotation
curve it produces, and attempt to reproduce a measured flat curve. Adding an
extended halo changes the force law used by the orbiting tracers and allows the
curve to remain flat at large radius, connecting the experiment with modern
rotation-curve data sets such as SPARC \citep{lelli2016}.

This activity illustrates a broader goal of \Gravitas: students should
experience why an inference was made. The conclusion that galaxies contain dark
matter becomes the resolution of a failed mass model rather than a fact
presented without its observational pressure.

\begin{figure*}[t]
\centering
\includegraphics[width=0.94\textwidth]{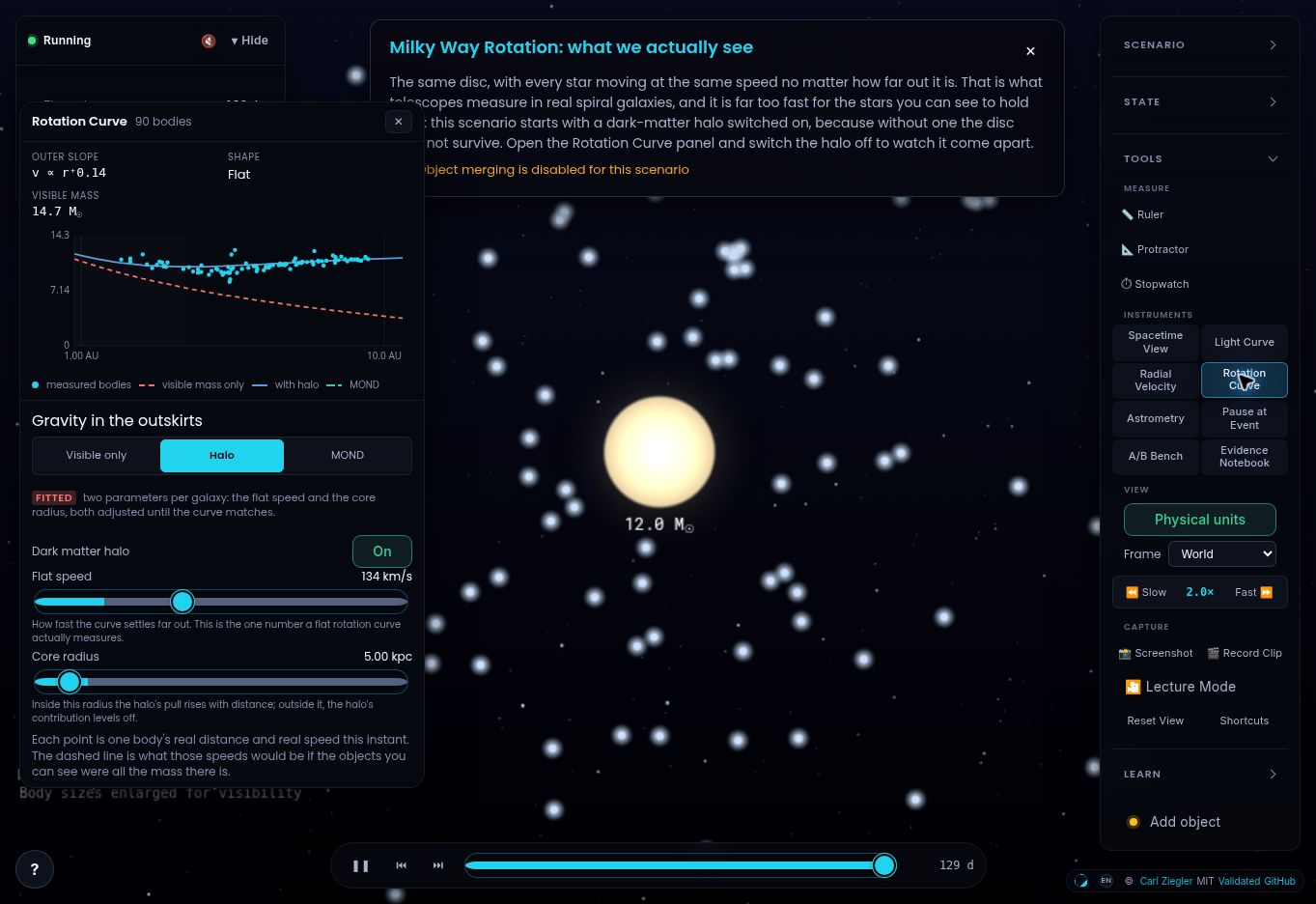}
\caption{The Missing Mass activity places the live galaxy and its rotation
curve in one workspace. Each cyan point is a simulated body's current distance
and speed; the dashed curve is the prediction from visible mass alone. Students
can compare visible-only, fitted-halo, and MOND options, change model parameters,
and watch the physical system respond rather than treating the curve as a
detached illustration.}
\label{fig:rotation-curve}
\end{figure*}

\section{Designed for Real Classrooms}
\label{sec:instructors}

An educational tool is only useful if it fits the time, equipment, and workflow
available to an instructor. \Gravitas\ supports several levels of use:

\begin{itemize}
\item \textbf{Live demonstration.} Lecture mode enlarges text, selects a
high-contrast daylight theme, supplies a spotlight pointer, and lets an
instructor step through a prepared sequence with arrow keys.
\item \textbf{Embedded figure.} A share link with embed mode produces a
chrome-free interactive view suitable for a course page or learning-management
system.
\item \textbf{Short activity.} Curated routes extract a coherent
prediction--experiment--explanation cycle from a longer investigation for use
during a single class meeting.
\item \textbf{Laboratory assignment.} A full investigation saves progress and
produces a student report that can be submitted through the course's existing
system.
\item \textbf{Open exploration or project.} The sandbox, A/B bench, share links,
and data export support questions designed by the learner or instructor.
\end{itemize}

For each guided investigation, the instructor collection includes learning
objectives, expected observations, common wrong turns, and an answer key
generated from the same lesson source used by students. An adopter's guide and
curriculum map help place activities into an existing course. Public teaching
pages explain the workflow and provide demonstration links before an instructor
requests access to protected answer materials.

\subsection{Adoption can begin with one class meeting}

An instructor does not need to redesign a course around the simulation. The
teaching collection presents the same scientific ideas in several formats, so a
first use can be a brief projected demonstration and a later use can become a
graded laboratory. Table~\ref{tab:classroom-patterns} shows typical entry
points. The durations are planning ranges rather than enforced timers; students
can pause, return, and restore a progress backup.

\begin{table*}[t]
\caption{Classroom-use patterns supported by the same simulation}
\label{tab:classroom-patterns}
\centering
\small
\renewcommand{\arraystretch}{1.15}
\begin{tabular}{llll}
\hline\hline
\TableCell{0.16\textwidth}{\textbf{Use}} &
\TableCell{0.13\textwidth}{\textbf{Typical time}} &
\TableCell{0.30\textwidth}{\textbf{Instructor preparation}} &
\TableCell{0.32\textwidth}{\textbf{Student evidence}}\\
\hline
\TableCell{0.16\textwidth}{Prediction demo} &
\TableCell{0.13\textwidth}{5--10 min} &
\TableCell{0.30\textwidth}{Open a prepared scenario or lecture sequence;
pose one question before pressing run.} &
\TableCell{0.32\textwidth}{Vote or short written prediction,
followed by a one-sentence revision.}\\
\TableCell{0.16\textwidth}{Focused activity} &
\TableCell{0.13\textwidth}{15--25 min} &
\TableCell{0.30\textwidth}{Share a prepared activity link with only the
controls and steps needed for the question.} &
\TableCell{0.32\textwidth}{One measurement, screenshot or
graph, and a claim supported by that evidence.}\\
\TableCell{0.16\textwidth}{Class-period investigation} &
\TableCell{0.13\textwidth}{35--50 min} &
\TableCell{0.30\textwidth}{Select a catalogue lesson by subject,
calculation load, and prerequisite.} &
\TableCell{0.32\textwidth}{Saved predictions, measurements, and an
exportable report.}\\
\TableCell{0.16\textwidth}{Multi-session laboratory} &
\TableCell{0.13\textwidth}{50--100 min} &
\TableCell{0.30\textwidth}{Assign a longer observing, stellar,
resonance, or gravitational-wave sequence.} &
\TableCell{0.32\textwidth}{Report PDF plus optional
unit-labelled CSV for plotting or fitting.}\\
\TableCell{0.16\textwidth}{Independent project} &
\TableCell{0.13\textwidth}{Several meetings} &
\TableCell{0.30\textwidth}{Provide a shared starting state and a
question, or ask students to propose both.} &
\TableCell{0.32\textwidth}{Reproducible state link, A/B
comparison, provenance file, and interpretation.}\\
\hline
\end{tabular}
\end{table*}

The most compact activities still preserve the essential cycle. For example,
an instructor can ask whether an eccentric planet moves fastest at periapsis or
apoapsis, collect predictions, run the scene, and use the event-aware stopwatch
or equal-area overlay to settle the question. A longer version asks students to
record multiple intervals, connect the result to angular momentum, and include
the measurements in a report. Because both versions use the same scenario, the
short activity is preparation for the lab rather than a separate demonstration
that must later be translated.

\subsection{Instructor-facing support}

The instructor materials are written around teachable decisions rather than
around interface features. Each guide states the learning objectives, estimated
time, prerequisite ideas, expected measurements, likely wrong turns, and a
suggested stopping point. Answer keys are generated from the same lesson data as
the student activity, reducing the risk that a changed question leaves an old
key behind. A curriculum map groups the investigations by concept and sequence;
the public teaching page offers six demonstrations that open directly in
reproducible states.

This division also protects the student experience. The model description,
validation record, demonstrations, and learning objectives are public. Detailed
answer material is distributed separately to instructors, while the simulation
itself remains a static site with no student login. The result is practical for
courses that already have an LMS: instructors continue to collect work where
they normally do, and \Gravitas\ supplies the experiment and the report.

Access considerations are treated as instructional infrastructure rather than
optional polish. The interface and all investigations are available in English
and Spanish. Keyboard focus management, screen-reader labels and a current text
summary of the canvas, reduced-motion styling, multiple themes, and automated
accessibility checks improve access. Important limitations remain: placing and
launching bodies has no keyboard equivalent, some curve shapes are not
narrated, and reduced-motion settings do not stop the orbital motion itself.
The application can operate offline after its assets have been cached,
including on lower-powered devices with a reduced visual-quality tier. Because
all computation and progress storage occur locally, students are not required
to create an account or send their work to a third-party service.

These features also make assignments more resilient. A link can reproduce the
starting state across a large class. A screenshot documents its own time and
scale. A CSV names the units in its columns. A report gathers the student's
predictions, measurements, graphs, and explanations. The goal is to reduce the
amount of custom technical scaffolding an instructor must build before the
physics can begin.

\section{Scientific Scope, Transparency, and Validation}
\label{sec:scope}

\Gravitas\ is an educational model, not a professional research code. Its core
dynamics are Newtonian and two-dimensional. Point-mass gravity is integrated in
a plane; close encounters use softening; and some scenarios allow selected
bodies to feel gravity without exerting it so that large tracer populations
remain responsive. Collisions are represented as perfectly inelastic mergers.
Relativistic quantities such as Schwarzschild radius may be evaluated
analytically, but general-relativistic orbital dynamics are not solved. Compact
binary decay in the open sandbox uses a phenomenological damping prescription;
the gravitational-wave lab instead uses a leading-order quadrupole inspiral,
terminates it at the Schwarzschild innermost stable circular orbit, and does not
model merger or ringdown. Visual elements such as accretion disks, jets, and
the optional rubber-sheet spacetime view are illustrative. Body sizes are
compressed for visibility and must not be measured from their drawn diameters.

These boundaries are stated prominently on the public model page. Features are
labelled as simulated, analytic, approximated, illustrative, or not modeled.
That vocabulary is also pedagogically useful: students can ask whether a
particular conclusion follows from the evolving equations, a closed-form
relationship, or a visual analogy. Instructors can decide whether the model is
appropriate for the learning objective before assigning it.

A public validation report currently contains \PhysicsChecks/\PhysicsChecks\
passing checks across \ValidationAreas\ areas and cites \ValidationSources\
sources. It covers orbital periods, Kepler relationships, conservation laws,
convergence behavior, escape speed, exoplanet observables, habitable-zone
quantities, rotation curves, black-hole scaling relations, and stored
parameters for real systems. The dashboard separates 87 analytic, 109
integrated, 34 published, 12 approximation, and one empirical check so that a
test of internal arithmetic is not presented as equivalent to agreement with
an observation. Each numeric tolerance has a stated justification, and the
entire suite can be rerun in the browser.

Scenario-stability checks audit the shipped catalogue, while browser tests
exercise the student workflows: opening instruments, changing an observer,
walking through a lesson, restoring a shared state, and generating a report.
These checks cannot establish that the software is error-free, nor can they
substitute for classroom evaluation. They do make the scientific claims,
tolerances, known approximations, and failure conditions inspectable.

\begin{figure*}[t]
\centering
\includegraphics[width=0.94\textwidth]{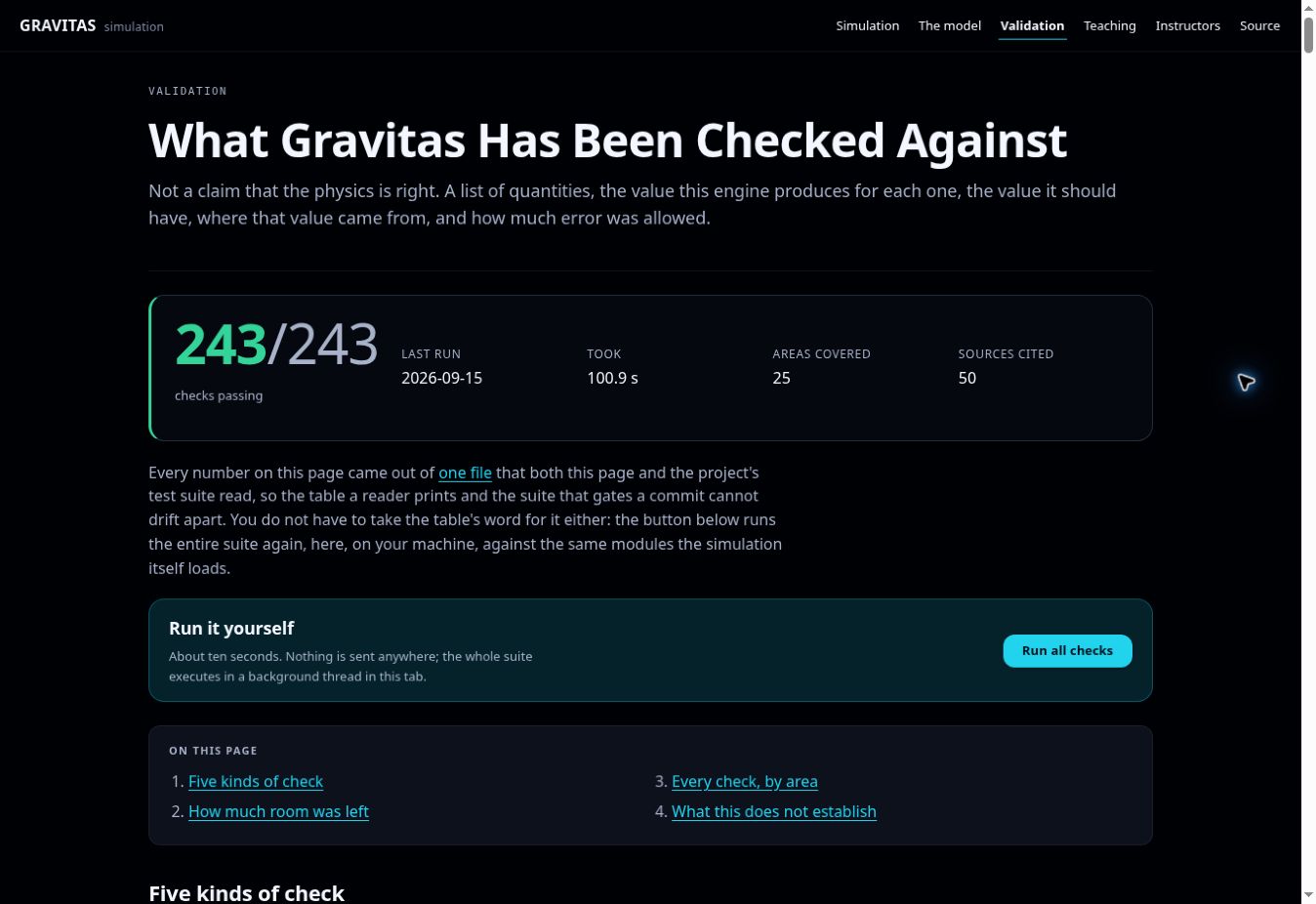}
\caption{The public validation dashboard reports the current result, date,
runtime, areas covered, and sources cited. It also explains what the checks do
and do not establish and lets a reader rerun the suite locally in the browser.
For an instructor, this provides a concise path from a classroom activity to
the evidence and limitations behind the model.}
\label{fig:validation}
\end{figure*}

Version \Version\ should therefore be evaluated along two separate dimensions.
Scientific verification asks whether the model produces the quantity it claims
to produce within an appropriate tolerance. Educational evaluation asks how
students and instructors use the environment and whether the intended
reasoning appears in their work. The first is included in the release process.
Formal evidence for the second is future work. The teaching site provides a
local feedback template so adopters can record course context, time on task,
friction points, and observed student reasoning without transmitting student
data.

\section{Evaluation and Responsible Use}
\label{sec:evaluation}

The version \Version\ paper makes a claim about capability, not yet a claim
about measured learning gains. The current software demonstrates that a student
can open the listed scenarios and investigations, collect the described
measurements, save progress, and export the stated products. The validation
suite tests whether the scientific quantities are produced as documented.
Neither result by itself shows that a particular course will achieve a larger
conceptual gain than another teaching method. That question requires classroom
data, and it should be evaluated separately from technical correctness.

A practical first evaluation phase is therefore adoption-focused. Instructors
can report whether a prepared activity fit the promised time, whether students
could begin without technical intervention, which steps produced useful
discussion, and which instructions interrupted the reasoning. Student work can
be sampled with a compact rubric: (1) prediction is explicit, (2) experiment
changes the intended variable, (3) evidence is relevant and labelled, (4)
explanation connects the evidence to the physical claim, and (5) limitation or
uncertainty is acknowledged when appropriate. The rubric applies across topics
and makes it possible to improve an investigation without reducing evaluation
to whether a student reached the expected number.

A later classroom study can pair selected investigations with short concept
questions before and after the activity, along with the evidence rubric and an
instructor implementation log. The catalogue's independent activities make
small pilots possible: an adopter can evaluate the retrograde-motion,
exoplanet, or rotation-curve activity without committing an entire course.
Where a comparison section is available, course context, time on task, and the
degree of instructor facilitation should be documented, since each can matter
as much as the presence of the simulation itself.

The privacy model simplifies that work but also sets a boundary. \Gravitas\ has
no student accounts and does not collect analytics or transmit investigation
answers. Formal studies must therefore obtain consent and collect data through
the institution's approved process rather than silently repurposing application
telemetry. For routine teaching, the same design means that a student can use
the full environment without surrendering personal data. Progress, backups,
reports, and CSV exports remain under the student's control.

For independent learners, responsible use includes knowing when to leave the
simulation and consult observations, derivations, or a higher-fidelity model.
The public model page names the approximations, the validation page names what
has been tested, and each specialized activity identifies whether a quantity is
integrated, analytic, phenomenological, or illustrative. The intended habit is
not to trust a polished picture automatically, but to ask what generated it and
what evidence would challenge it.

\section{Availability and Reuse}
\label{sec:availability}

The simulation is available at \url{https://gravitas-sim.online}. Source code,
documentation, model limitations, validation materials, and issue tracking are
available at
\url{https://github.com/gravitas-sim/gravitas-sim.github.io}. The project is
released under the MIT License.

\noindent\begin{minipage}{\columnwidth}
The source archive for version \Version\ is preserved by Zenodo:

\begin{center}
\textbf{DOI:} \href{https://doi.org/\ReleaseDOI}{\nolinkurl{\ReleaseDOI}}\\
\textbf{Zenodo record:} \url{\ZenodoRecord}
\end{center}

The archive corresponds to the GitHub \texttt{v\Version} release at source
commit \texttt{\CodeSnapshot}. The version DOI above identifies this immutable
release rather than the evolving project as a whole.
\end{minipage}

Instructors may use the built-in investigations as written, assign selected
steps, or create their own activities from shared scenarios and exported data.
The repository welcomes reports from classroom use, translations, new
scenarios, and proposed investigations. The most valuable feedback is concrete:
what learners predicted, what they measured, where an interface choice
interrupted the reasoning, and what explanation emerged after the experiment.

\section{Conclusion}
\label{sec:conclusion}

\Gravitas\ gives students a laboratory for systems that ordinary laboratories
cannot hold. A planet can be moved, a viewpoint changed, a dark halo removed, a
binary replayed from the same initial state, or an observing schedule redesigned
within minutes. More importantly, the environment asks students to make those
changes for a reason: to test a prediction and construct an explanation from
measured evidence.

For an instructor, the entry cost is deliberately low: open a link, choose a
scenario or investigation, and use it as a projected demonstration, an embedded
interactive figure, a short activity, or a complete lab. For a student, the
invitation is equally direct: build a system, run it, measure it, and find out
whether the universe in the model behaves as expected. \Gravitas\ is available
now at \url{https://gravitas-sim.online}, and the versioned \Version\ archive
is citable at DOI \doi{\ReleaseDOI}.

\section*{Acknowledgments}
\Gravitas\ began as a Summer Undergraduate Research Experience (SURE) project
at Stephen F. Austin State University and was funded by the SFA College of
Sciences and Mathematics. The author welcomes feedback from instructors and
students who use \Gravitas\ in a course.

\software{\Gravitas\ \citep{gravitas_zenodo}}

\bibliography{references}

@misc{gravitas_zenodo,
  author       = {Ziegler, Carl},
  title        = {{Gravitas}: An Interactive Astrophysics Sandbox for Teaching},
  year         = {2026},
  month        = sep,
  version      = {1.0.0},
  publisher    = {Zenodo},
  doi          = {10.5281/zenodo.22800610},
  url          = {https://doi.org/10.5281/zenodo.22800610}
}

@article{finkelstein2005,
  author  = {Finkelstein, Noah D. and Adams, Wendy K. and Keller, Christopher J.
             and Kohl, Patrick B. and Perkins, Katherine K. and Podolefsky,
             Noah S. and Reid, Sam and LeMaster, Ronald},
  title   = {When Learning about the Real World Is Better Done Virtually:
             A Study of Substituting Computer Simulations for Laboratory
             Equipment},
  journal = {Physical Review Special Topics--Physics Education Research},
  year    = {2005},
  volume  = {1},
  number  = {1},
  pages   = {010103},
  doi     = {10.1103/PhysRevSTPER.1.010103}
}

@article{wieman2008,
  author  = {Wieman, Carl E. and Adams, Wendy K. and Perkins, Katherine K.},
  title   = {{PhET}: Simulations That Enhance Learning},
  journal = {Science},
  year    = {2008},
  volume  = {322},
  number  = {5902},
  pages   = {682--683},
  doi     = {10.1126/science.1161948}
}

@article{rutten2012,
  author  = {Rutten, Nico and van Joolingen, Wouter R. and van der Veen,
             Jan T.},
  title   = {The Learning Effects of Computer Simulations in Science Education},
  journal = {Computers \& Education},
  year    = {2012},
  volume  = {58},
  number  = {1},
  pages   = {136--153},
  doi     = {10.1016/j.compedu.2011.07.017}
}

@article{hake1998,
  author  = {Hake, Richard R.},
  title   = {Interactive-Engagement versus Traditional Methods: A Six-Thousand-
             Student Survey of Mechanics Test Data for Introductory Physics
             Courses},
  journal = {American Journal of Physics},
  year    = {1998},
  volume  = {66},
  number  = {1},
  pages   = {64--74},
  doi     = {10.1119/1.18809}
}

@article{freeman2014,
  author  = {Freeman, Scott and Eddy, Sarah L. and McDonough, Miles and Smith,
             Michelle K. and Okoroafor, Nnadozie and Jordt, Hannah and
             Wenderoth, Mary Pat},
  title   = {Active Learning Increases Student Performance in Science,
             Engineering, and Mathematics},
  journal = {Proceedings of the National Academy of Sciences},
  year    = {2014},
  volume  = {111},
  number  = {23},
  pages   = {8410--8415},
  doi     = {10.1073/pnas.1319030111}
}

@article{abbott2016,
  author  = {{Abbott}, B. P. and others},
  collaboration = {LIGO Scientific Collaboration and Virgo Collaboration},
  title   = {Observation of Gravitational Waves from a Binary Black Hole
             Merger},
  journal = {Physical Review Letters},
  year    = {2016},
  volume  = {116},
  number  = {6},
  pages   = {061102},
  doi     = {10.1103/PhysRevLett.116.061102}
}

@article{dotter2016,
  author  = {Dotter, Aaron},
  title   = {{MESA} Isochrones and Stellar Tracks ({MIST}) 0: Methods for the
             Construction of Stellar Isochrones},
  journal = {The Astrophysical Journal Supplement Series},
  year    = {2016},
  volume  = {222},
  number  = {1},
  pages   = {8},
  doi     = {10.3847/0067-0049/222/1/8}
}

@article{choi2016,
  author  = {Choi, Jieun and Dotter, Aaron and Conroy, Charlie and Cantiello,
             Matteo and Paxton, Bill and Johnson, Benjamin D.},
  title   = {{MESA} Isochrones and Stellar Tracks ({MIST}). I. Solar-Scaled
             Models},
  journal = {The Astrophysical Journal},
  year    = {2016},
  volume  = {823},
  number  = {2},
  pages   = {102},
  doi     = {10.3847/0004-637X/823/2/102}
}

@article{charbonneau2000,
  author  = {Charbonneau, David and Brown, Timothy M. and Latham, David W. and
             Mayor, Michel},
  title   = {Detection of Planetary Transits across a Sun-like Star},
  journal = {The Astrophysical Journal},
  year    = {2000},
  volume  = {529},
  number  = {1},
  pages   = {L45--L48},
  doi     = {10.1086/312457}
}

@article{kopparapu2013,
  author  = {Kopparapu, Ravi Kumar and Ramirez, Ramses and Kasting, James F. and
             Eymet, Vincent and Robinson, Tyler D. and Mahadevan, Suvrath and
             Terrien, Ryan C. and Domagal-Goldman, Shawn and Meadows, Victoria
             and Deshpande, Rohit},
  title   = {Habitable Zones around Main-Sequence Stars: New Estimates},
  journal = {The Astrophysical Journal},
  year    = {2013},
  volume  = {765},
  number  = {2},
  pages   = {131},
  doi     = {10.1088/0004-637X/765/2/131}
}

@article{lelli2016,
  author  = {Lelli, Federico and McGaugh, Stacy S. and Schombert, James M.},
  title   = {{SPARC}: Mass Models for 175 Disk Galaxies with
             Spitzer Photometry and Accurate Rotation Curves},
  journal = {The Astronomical Journal},
  year    = {2016},
  volume  = {152},
  number  = {6},
  pages   = {157},
  doi     = {10.3847/0004-6256/152/6/157}
}
\bibliographystyle{aasjournal}

\end{document}